\documentclass[11pt, hidelinks]{article}

\usepackage[colorlinks=true, linkcolor=black, urlcolor=blue, citecolor=blue, anchorcolor=blue]{hyperref}
\usepackage{wrapfig}
\usepackage{comment}
\usepackage{textgreek}
\usepackage{url}
\usepackage{fullpage}
\usepackage[titletoc,title]{appendix}
\usepackage{enumitem, comment, xifthen}
\let\mathbb\mathbf
\usepackage{amsmath,amssymb,amsthm, amsfonts}
\usepackage[T1]{fontenc}
\usepackage{makecell}
\usepackage{booktabs}

\usepackage{subcaption}
\usepackage[dvipsnames]{xcolor}
\DeclareMathOperator*{\E}{\bf E}
\usepackage[natbibapa]{apacite}
\usepackage{natbib}
\makeatletter
\long\def\@makecaption#1#2{
        \vskip 0.8ex
        \setbox\@tempboxa\hbox{\small {\bf #1.} #2}
        \parindent 1.5em 
        \dimen0=\hsize
        \advance\dimen0 by -3em
        \ifdim \wd\@tempboxa >\dimen0
                \hbox to \hsize{
                        \parindent 0em
                        \hfil 
                        \parbox{\dimen0}{\def\baselinestretch{0.96}\small
                                {\bf #1.} #2
                                } 
                        \hfil}
        \else \hbox to \hsize{\hfil \box\@tempboxa \hfil}
        \fi
        }
\makeatother
\usepackage{booktabs}
\usepackage{hhline}
\usepackage{array,multirow}
\usepackage{lipsum}
\usepackage{siunitx,etoolbox}

\usepackage{graphicx}
\graphicspath{{figs/}}

\begin{document}
\newcommand{\defeq}{:=}
\newcommand{\bftab}{\fontseries{b}\selectfont}
\newcommand{\tonset}{\ensuremath{{t_{\rm on}}}}
\newcommand{\naive}{\ensuremath{E}_{\rm naive}}
\newcommand{\observed}{\ensuremath{E}_{\rm obs}}
\newcommand{\recprob}{\ensuremath{\rho^{(1)}_{\tonset, g}}}
\newcommand{\deathprob}{\ensuremath{\rho^{(2)}_{\tonset, g}}}
\newcommand{\undiagprob}{\ensuremath{\rho^{(3)}_{\tonset, g}}}
\newcommand{\recnum}{\ensuremath{N^{(1)}_{\tonset, g}}}
\newcommand{\deathnum}{\ensuremath{N^{(2)}_{\tonset, g}}}
\newcommand{\undiagnum}{\ensuremath{N^{(3)}_{\tonset, g}}}
\newcommand{\totalnum}{\ensuremath{N^\ast_{\tonset, g}}}
\newcommand{\simind}{\ensuremath{\overset{\rm ind.}{\sim}}}
\newcommand{\deathavg}{\ensuremath{t_{\rm avg}}}
\newcommand{\rpcomment}[1]{\textcolor{red}{RP---#1}}
\newcommand{\final}{\ensuremath{{T}}}
\newcommand{\e}{\ensuremath{\mathrm{e}}}
\newcommand{\R}{\ensuremath{\mathbf{R}}}
\newcommand{\edge}[2]{\mathrm{#1}\protect\to\mathrm{#2}}
\newcommand{\upstairs}[1]{\textsuperscript{#1}}
\newcommand{\affilone}{\dag}
\newcommand{\affiltwo}{\ddag}
\newcommand{\affilthree}{$\diamond$}
\newcommand{\eg}{\textit{e}.\textit{g}.,}
\renewcommand{\cite}[1]{\citep{#1}}

\let\star\ast
\let\epsilon\varepsilon
\let\phi\varphi

\renewcommand\theadfont{\normalsize}
\begin{center}

  {\bf{\Large On Identifying and Mitigating Bias in the Estimation\\[.2cm] of the COVID-19 Case Fatality Rate}} \\

  \vspace*{.2in}
  
  \begin{tabular}{cc}
    Anastasios Nikolas Angelopoulos\upstairs{\affilone,*}, Reese Pathak\upstairs{\affilone},
    Rohit Varma\upstairs{\affilthree}, and Michael I.\ Jordan\upstairs{\affilone, \affiltwo,*}\\[0.25ex]
   {\small \upstairs{\affilone} Department of Electrical Engineering and Computer Science, UC Berkeley} \\
   {\small \upstairs{\affilthree} Southern California Eye Institute, CHA Hollywood Presbyterian Medical Center, Los Angeles} \\
   {\small \upstairs{\affiltwo} Department of Statistics, UC Berkeley} \\
   \texttt{\string{angelopoulos, pathakr,jordan\string}@cs.berkeley.edu}, \texttt{rvarma@sceyes.org} \\  
   {\small * corresponding authors}
  \end{tabular}
  
  \vspace*{0in}

\begin{abstract}
\small
The relative case fatality rates (CFRs) between groups and countries are key measures of relative risk that
guide policy decisions regarding scarce medical resource allocation during the ongoing COVID-19 pandemic.
In the middle of an active outbreak when surveillance data is the primary source of information,
estimating these quantities involves compensating for competing biases in time series of deaths, cases, and recoveries. These include time- and severity- dependent reporting of cases as well as time lags in observed patient outcomes.
In the context of COVID-19 CFR estimation, we survey such biases and their potential significance. 
Further, we analyze theoretically the effect of certain biases, like preferential reporting of fatal cases, on naive estimators of CFR. We provide a partially corrected estimator of these naive estimates that accounts for time lag and imperfect reporting of deaths and recoveries. We show that collection of randomized data by testing the contacts of infectious individuals regardless of the presence of symptoms would mitigate bias by limiting the covariance between diagnosis and death. 
Our analysis is supplemented by theoretical and numerical results and a simple and fast  
\href{https://github.com/aangelopoulos/cfr-covid-19}{open-source codebase}.\footnote{\url{https://github.com/aangelopoulos/cfr-covid-19}}
\end{abstract}
\end{center}

\section{Introduction}
\label{sec1}
As of May 18, 2020, the 2019 novel Coronavirus (SARS-CoV-2) outbreak has claimed at least $317,000$ lives out 
of $4.8$ million confirmed cases worldwide, of which $1.8$ million recovered~\cite{jhu-csse}. Because the basic 
reproduction number $R_0$ of the virus is high (estimated to fall between $2$ and $3$ by \citealt{liu2020reproductive}),
public health organizations  and local, state, and national governments must allocate scarce resources to populations especially susceptible to death during this pandemic. Therefore it is critical to have good estimates of the proportion of fatal infections of COVID-19: this quantity is referred to as the absolute case fatality rate (CFR).\footnote{An important caveat at the outset: As Box 1 of \citet{lipsitchsummary} points out, ``The CFR\dots is an ambiguous term, as its definition and value depend on what qualifies an individual to be a {`case.'}'' In this article, we are defining the CFR as the proportion of fatal infections; that is, the proportion of deaths among all COVID-19 infected individuals.  This is a version of the CFR that is often called the infection fatality rate (IFR). As we will discuss, even this definition is ambiguous for many reasons, including the cause of death. While no perfect definition exists, due in part to the many biases described in Section~\ref{sec:biases}, in any
given analysis it will be important to choose a pertinent definition, and to take care when making comparisons between analyses that have chosen different definitions.}
It is additionally important to understand the relative CFRs between
different subpopulations (i.e., the ratio of their absolute CFRs). We view the
relative CFR as a useful target for data-informed resource-allocation protocols
because it is a key measure of relative risk.  Indeed, the absolute CFR is a
measure of absolute severity only for a particular population, since it averages
out effects of medical care, age, geography, genetics, and more.
Practical decisions will ultimately be made based on coarse stratifications of these covariates; for example,
a relative CFR that is specific to a geographical region may be needed for resource
planning and allocation. Similarly, a relative CFR that is specific to sex or race is often sought
to monitor for and ensure equitable treatment of patients within hospitals across demographics. To
facilitate such planning, we target the relative number of deaths among total cases
between groups of people (e.g., senior citizens in Italy) as a critical measure of
relative risk that informs decisions affecting human lives. Other such measures include prevalence and risk of hospitalization.

It is widely believed that the naive estimator of CFR, $\naive$, obtained from a simple ratio of reported deaths to 
reported cases (and which has a value of $6.6\%$ when applied to the data of May 18, 2020), is 
biased~\cite{fauci,caution}. Indeed, an extensive epidemiological literature has asserted this
bias and presented methods that attempt to mitigate it~\cite{sars20yr}.  Bias-mitigation methods are also present in a
large literature on survey sampling and weighting~\cite{gelmanrejoinder}. Despite this academic
background, naive estimates continue to be used, reported, and cited in major 
publications~\cite{chinataskforce,jamacfr,lipsitchroast}. 

Since publicly available health surveillance data for COVID-19 are heterogeneous and partially observed,
it is problematic to assert that any estimator uniformly outperforms the naive
estimator. A variety of competing (and unknown) biases,
both negative and positive, could conceivably cancel,
causing the naive estimator to be closer to the true CFR despite its theoretical inadequacy.
Statistical wisdom would suggest that the conundrum of conflicting biases can be remedied by
studying the multi-stage process that relates data obtained by surveillance sampling to the populations that are the target of inferential assertions.  
It is the goal of this article to present 
such a statistical perspective and explore some of its consequences for COVID-19.

Clarity on the potential biases underlying the use of data from
surveillance sampling can help to determine what additional datasets
may be needed to mitigate bias. Examples that will inform
our discussion include the New York seroprevalence study reported in \citet{nycprevalence}, which helped correct significant under-ascertainment of mild cases,
and~\citet{fergusoncfr}, who made use of individualized
case data, polymerase-chain reaction (PCR) prevalence data, and Bayesian inferential
methods, resulting in a CFR estimate of $1.38\%$. 
These studies can improve public-health response to COVID-19 insofar as they are
accompanied by an understanding of their implicit assumptions, including putative control of possible biases. 

The remainder of this article is organized as follows. In Section~\ref{sec:biases}, we provide a statistical perspective
on the many potential biases affecting absolute and relative CFR estimation.
In Section~\ref{sec:naive}, we employ the general perspective to isolate some restricted contexts in which
two naive estimators are unbiased, with implications on the need for contact tracing.
In Section~\ref{sec:models}, we consider how model-based inference can expand the contexts
in which unbiased estimation is possible.  We provide an
illustrative example, showing how an (approximate) maximum likelihood estimator
from~\citet{reich} can be applied to correct bias from relative
reporting rates of fatal and resolved cases using only surveillance data and
an approximate horizon distribution of deaths. We discuss how the general principle
of coping with incomplete data via Poisson approximation and a log-linear likelihood model can be more widely applied. In Section~\ref{sec:results}, we present results of this method on COVID-19 data. Finally, in Section~\ref{sec:discussion}, we give a mathematical justification for contact tracing as a data-collection methodology~\cite{contactbook,chowellbook}, and discuss how it would mitigate many of the problematic biases at their source.

\section{Sources of Bias in COVID-19 Surveillance Data}
\label{sec:biases}
\begin{figure}
    \centering
    \begin{subfigure}{\textwidth}
        \vspace{-7mm}
        \centering
        \includegraphics[width=0.75\textwidth]{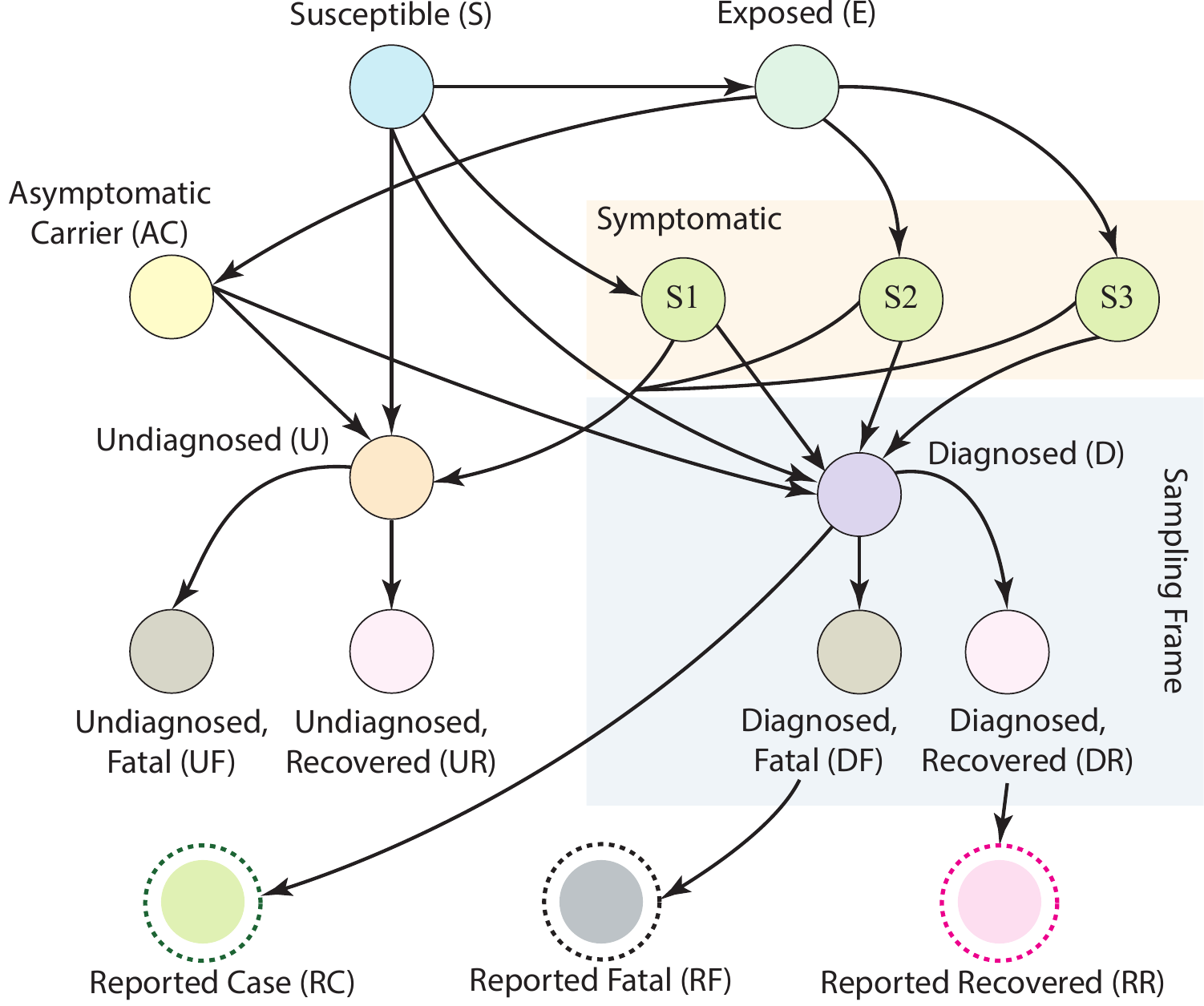}
    \end{subfigure}
    \begin{subfigure}{\textwidth}
        \vspace{1mm}
        \centering
        \small
        \begin{tabular}{ @{}l p{0.785\textwidth}@{} }
            \textbf{Edge} & \textbf{Notes and some potential sources of sampling and estimation bias} \\
            \midrule
            S $\to$ E & Social distancing, occupation, family size, behavior.\\
            S $\to$ U & COVID-negative people who are not diagnosed. \\ 
            S $\to$ D & Assay specificity and sensitivity, test availability. \\
            S $\to$ S1 & Group characteristics such as genetics and immunity. Exposure to flu.\\
            E $\to$ AC,S2 & Infectious dose, route of transmission, and group characteristics. \\
            E $\to$ S3 & Presence of other underlying medical conditions. \\
            AC $\to$ U,D & Random sampling, contact tracing, test availability.\\
            S1 $\to$ U & Assay specificity and sensitivity, case severity. \\
            S2 $\to$ U & Assay specificity and sensitivity, case severity, test availability.\\
            S3 $\to$ U & Misattribution of symptoms, assay specificity and sensitivity, comorbidities.\\
            S1 $\to$ D & Misattribution, assay specificity and sensitivity, group characteristics.\\
            S2 $\to$ D & Delays in seeking care, interventions like contact tracing, group characteristics. \\
            S3 $\to$ D & Assay specificity and sensitivity, contact tracing and test availability.\\
            U $\to$ UF,UR & Group characteristics, particularly comorbidities.\\
            D $\to$ DF & Misattribution, group characteristics, imperfect reporting.\\
            D $\to$ DR & Survey nonresponse, group characteristics, imperfect reporting.\\
            D $\to$ RC & Imperfect reporting: errors, case definition, release of incorrect data, time lag. \\
            DF $\to$ RF &  Imperfect reporting, national reporting guidelines, ease of reporting.\\
            DR $\to$ RR &  Imperfect reporting, national reporting guidelines, ease of reporting.\\
        \end{tabular}
    \end{subfigure}
    \caption{\small \textbf{Some sources of bias arising from COVID-19 health surveillance data.}
      For simplicity, nodes in this diagram can be thought of as representing the number of people in the labeled state. Each edge represents a conditional probability of transitioning between states and has an associated time lag. S1 represents COVID-19 negative people with flu symptoms. S2 represents people with symptoms \textit{caused} by COVID-19. S3 represents COVID-19 positive people with symptoms caused by another underlying health condition. The sampling frame of population surveillance data based on standard hospital reporting is in light blue; values of edges outside the sampling frame cannot be inferred only with data from within the sampling frame. Relative time lag introduces bias across all edges.} 
    \label{graphicalmodel}
\end{figure}
In Figure~\ref{graphicalmodel}, we present a graphical model that captures aspects of the data-generating process for COVID-19 
surveillance data.  Graphical models provide a general formal language for reasoning about the probabilistic and causal structure of collections of random variables (see, e.g., \citealt{jordan2004}).  In this article, it suffices to think of the model in Figure~\ref{graphicalmodel} informally as a depiction of dependencies among population-level and sampled quantities in surveillance data.  Our objective is to consider biases that may arise along each edge of the graph. Prior work on SARS, H1N1, H7N9, H5N1, MERS, and HIV has identified or even quantified many of these biases (\citealt{gr1,gr2}; see \citealt{lipsitchsummary} for a review). 
The diagram also depicts the population eligible to be sampled (the `sampling frame') through collection of surveillance data from standard hospital reports and death certificates. Asymptomatic carriers are excluded from the sampling frame because testing is currently not recommended or available for asymptomatic individuals. We roughly categorize biases as: under-ascertainment of mild cases, time lags, interventions, group characteristics, and imperfect reporting and attribution. Extensive (but not comprehensive) discussion of the magnitude and direction of these biases corroborated by COVID-19-specific evidence is included in the following subsections. 

These competing biases can often be expressed in terms of ratios of edge weights (conditional probabilities) from Figure~\ref{graphicalmodel}. For example, with $\edge{A}{B}$ denoting the value of the edge between nodes A and B, if the probability of reporting a fatal case is greater than reporting an infection ($\edge{DF}{RF}>\edge{D}{RC}$), then ignoring other biases, $\naive$ will be upwardly biased by a factor $b>1$. However, because biases can compete with one another, this does not mean $(1/b)\naive$ is always a better estimator than $\naive$. Speaking loosely, the bias incurred by under-ascertainment of asymptomatic cases could be $1/b$, canceling out the former bias. Accordingly, the total error of any estimator is indeterminate. Therefore, the use of estimators based on surveillance data requires being clear on underlying assumptions.  Statisticians and public health officials should proceed with multiple estimation strategies, armed with an understanding of the accompanying biases, and they should endeavor to collect additional data that mitigates the biases (as we describe in Section~\ref{sec:discussion}).

Figure~\ref{graphicalmodel} conveys two important takeaways regarding the estimation of CFR. First and foremost, information about edges outside the sampling frame cannot be inferred from data within the sampling frame alone. Even within the frame, data is compromised by the biases listed in the subsections below. Each incoming edge to D can bias CFR estimates up or down, depending on its ratio to other incoming edges. This cannot be disentangled only by looking at the value of D. Secondly, even within the sampling frame, the relative values and time lags of $\edge{D}{DF}$ and $\edge{D}{DR}$ affect the estimation of CFR. However, these edges may be the only ones subject to correction using population-level surveillance data alone. This motivates our choice of an illustrative estimator of relative CFR, adapted from \citet{reich}. In particular, the estimator is based on assumptions under which estimation of relative CFR is possible while correcting for the relative values and time lags of $\edge{D}{DF}$ and $\edge{D}{DR}$.

In the following subsections, we will cite evidence for the existence, magnitudes, and directions of certain biases from Figure~\ref{graphicalmodel}. Our analysis was done in mid-April 2020. We categorize biases resulting from one of five phenomena: under-ascertainment of mild cases, time lags, interventions, group characteristics, and imperfect reporting and attribution.

\subsection{Under-ascertainment of Mild Cases}
Diagnosing severe cases more often than mild cases will falsely increase CFR. In Figure~\ref{graphicalmodel}, this bias corresponds most directly to spuriously increasing $\edge{AC}{U}$ and $\edge{S2}{U}$ and/or decreasing $\edge{AC}{D}$ and $\edge{S2}{U}$. This bias may have high magnitude, since the true number of infections is likely to be several times as high as the reported number of cases in countries where testing is limited~\cite{fauci}. The significantly lower CFR in South Korea, a country with widespread testing, corroborates this explanation~\cite{jhu-csse}. A recent serology study from New York City (reported in \citealt{nycprevalence}) suggests the prevalence of COVID-19 is 21.2\%, much higher than the confirmed case count. This indicates that the number of infections may be much larger than surveillance data implies globally. 

\subsection{Time Lags}
Deaths and recoveries are reported after cases are confirmed, which artificially decreases the naive CFR~\cite{wilsonearly}. More specifically, when a number of new cases is reported without delay, $\naive$ becomes biased downward since the deaths yet to occur from the new cases will be missing from the numerator. In Figure~\ref{graphicalmodel}, this means a time lag is incurred across edges $\edge{D}{DF}$ and $\edge{D}{DR}$. The value of this time lag depends on how early on in the disease's course a patient is diagnosed. The median value of the lag across $\edge{D}{DF}$ was estimated by~\citet{gamma1} in China in early February to be 6.7 days (Lognormal 95\% CI: $5.3-8.3$). Tracking individual cases and including them only after death or recovery would solve this problem. However, data on individuals is rare, as hospitals and/or governments generally report population-level data only. 

Fortunately, these edges are within the sampling frame, so we can have some hope of correcting the bias incurred by this time lag. In Section~\ref{sec:models}, we implement an estimator that handles this correction directly, and discuss the many further assumptions that must be made to assure its validity even within the sampling frame. In reality however, time lags affect every edge in Figure~\ref{graphicalmodel}. The `incubation period,' for example, creates lag across edges $\edge{E}{AC,S1,S2,\textrm{ and }S3}$, and was estimated at median 4.3 days based on~\citet{gamma1}. The time between onset and hospital admission, which is not perfectly represented by the graphical model, creates lag across edges $\edge{S1,S2,S3}{D}$ and had an estimated median of 1.5 days among recovered cases and 5.1 days among fatal ones~\cite{gamma1}. The large discrepancy between hospitalization lags of patients with different outcomes suggests the presence of an unknown bias factor. For example, earlier hospitalization may result in more effective treatment, canceling out the propensity of severe cases to seek care more quickly in the data collected by Linton et al. A complete discussion of the effects of all time lags across edges in Figure~\ref{graphicalmodel} is outside the scope of this article.

\subsection{Interventions}
Data collected after a recent government intervention targeted to lower transmission of COVID-19 could produce a spuriously increased CFR. One primary tool of governments is the imposition of social-distancing measures, which decrease the amount and initial dose of exposures, thereby lowering $\edge{S}{E}$. One incubation period after such a measure is enacted, the number of new cases will decrease, but the number of new deaths will not, since these deaths will be from cases diagnosed before the government intervention. This will upwardly bias the CFR for a few weeks after the intervention. 

As in other pandemic influenzas, there may be a direct biological effect of increasing the infectious dose, leading to higher fatality rates~\cite{infectiousdose}. Given an effective government intervention like social distancing, the infectious dose would decrease, directly lowering $\edge{U}{UF}$ and $\edge{D}{DF}$ and increasing $\edge{U}{UR}$ and $\edge{D}{DR}$. This will upwardly bias current CFR estimates for some weeks after the initial intervention, since new deaths will still occur from cases whose onset time was before the intervention. The Centre for Evidence Based Medicine has a helpful page dedicated to COVID-19 viral dynamics like these~\cite{cebm}.

Interventions to improve the quality of medical care can cause a drastic decrease in CFR, particularly when treatment options (e.g., drugs, blood transfusions, ventilators) become available or if training of health care workers (HCWs) improves. The effect can be highly pronounced in developing countries~\cite{choleratraining} and is the subject of active study today~\cite{estherinterventions,ebert}. By the same logic as above, these interventions can lead to a spuriously increased CFR estimate. However, interventions that improve accessibility of medical care, such as the new health facilities being constructed around the world~\cite{newhospitals1,newhospitals2}, can also increase testing and reporting. This increase will likely result in better data in the long term due to higher ascertainment of mild cases, although we have no data to support this conjecture.

\subsection{Group Characteristics}
It is already well understood that certain groups have a higher CFR than other groups. In other words, the edges $\edge{D}{DF}$, $\edge{D}{DR}$, $\edge{U}{UF}$, and $\edge{U}{UR}$ will have different values based on the characteristics of the sampled population, which could cause bias in either direction when estimating CFR. For example, the risk of death may be 34 to 73 times lower in people under 65 years old compared to those over 65~\cite{ioannidisage}. Furthermore, the incidences of comorbidities such as obesity, heart disease, smoking, genetics, and diabetes correlate with nation, socioeconomic status, race, sex, and more~\cite{smoking,obesitypoverty,heartafrica}. In the context of surveillance data, without knowing the proportion of these groups in the sampling frame, which may not be uniform in time, the CFR can be biased in either direction. \citet{stanfordcounty} argue that incorporating county-level data about these covariates can result in a more equitable public-health response.

\subsection{Imperfect Reporting and Attribution}
Both the definition of a `case' and also the criteria under which an individual is eligible for testing can bias CFR estimates. Case definition, even within one nation, can change case counts dramatically. On February 12, for example, the Chinese government changed the definition of `confirmed case' to include symptom-based diagnoses, resulting in a $600\%$ increase in cases that day~\cite{chinacase}. Without information on how deaths were attributed beforehand, we do not know the magnitude of this bias. Serious problems have been introduced by poor reporting on behalf of governments. For example, the Johns Hopkins GitHub stopped providing surveillance data on recovered cases within the United States, because the quality of the data was too low~\cite{jhuissue}. Furthermore, because testing is often reserved for the most severe cases, $\edge{S2}{D}$ is inflated while $\edge{AC}{D}$ is deflated~\cite{selectivetesting}. This will spuriously increase $\naive$.  Evidently, detailed knowledge of how cases, deaths, and recoveries are defined and reported are prerequisite to understanding these biases, even if it will be impossible to correct for them without finer-grained data. 

Sensitivity and specificity of COVID-19 tests certainly affect all of $\edge{AC,S1,S2,S3}{D}$ and $\edge{AC,S1,S2,S3}{U}$. A diagnostic test with a high false discovery rate will increase $\edge{S}{S1}$, incorrectly inflating the denominator of $\naive$ and spuriously decreasing the CFR. Nonetheless, assays have improved with time; the initial test developed by the U.S. Centers for Disease Control was ineffective~\cite{cdctest}. Still, the serology assay used by \citet{Bendavid2020} had a putative sensitivity of 80\% and specificity of 99.5\%, which may still be too low to provide estimates of a small prevalence.

Distinctly from under-ascertainment, reporting of infectious disease by health care providers in the United States is often incomplete and normally has a mean time delay of 12 to 40 days depending on the pathogen~\cite{reportingtime}. This means edges $\edge{D}{RC}$, $\edge{DF}{RF}$, and $\edge{DR}{RR}$ are not $1.0$. Because deaths are more likely to be reported by health care providers than confirmed cases or recoveries, ignoring time delay, $\edge{D}{RC}$ is less than $\edge{DF}{RF}$, biasing $\naive$ upward. Depending on the relative time delays across these edges, estimators may be biased in either direction. For example, if $\edge{D}{RC}$ is lagged more than $\edge{D}{RD}$, it would bias $\naive$ downward during the growth phase of an epidemic. To our knowledge, these time delays, which occur on a hospital-by-hospital basis, have not been quantified, and it is not obvious in what direction they will skew. The magnitude of this bias could be quite large for COVID-19. On Friday, April 17, the Wuhan government reported 1,290 new fatalities, increasing their cumulative death toll by 50\% in one day. They claimed the revision was because ``medical workers \dots might have been preoccupied with saving lives, and there existed delayed reporting, underreporting, or misreporting''~\cite{wuhandata}. This is a salient example of a high-magnitude bias from reporting errors that we can not correct, since we do not know at what time those deaths truly occurred.

Evidence from past epidemics also indicates this bias may be significant for COVID-19. Even for severe illnesses such as Hepatitis C, health care providers can underreport cases by up to 12x~\cite{hepc}. Historically, the magnitude of underreporting depends heavily on ease of reporting for HCWs (e.g., electronic vs. paper systems) and also mandatory reporting laws~\cite{reportingmandatory,reportingelectronic}. 

Finally, although survivorship bias may be small, misattribution of deaths (i.e., increased weight of S3 $\to$ D) may be significant. A recent JAMA article argued that COVID-19 positive patients with cardiac injury have a relative risk of death of 4.26 compared to those with no cardiac injury. Most of those patients also had abnormal electrocardiograms~\cite{heartcovid}. Another case study described a healthy 53-year-old woman who tested positive for COVID-19, did not show any respiratory involvement, but developed acute myopericarditis with systolic dysfunction~\cite{heartcovid2}. Kidney involvement has also been found~\cite{kidneycovid}. It is unclear how deaths in the presence of multiple diagnoses are being counted, and indeed, to which disease they should be attributed. Disentangling these relationships may be possible with regression on high-resolution clinical data. However, we have not seen this level of detail reflected in surveillance data. Comparisons with historical mortality data suggest tens of thousands of deaths are misattributed or unreported~\cite{nythistoricaldeaths}.

\section{Naive Estimators}
\label{sec:naive}
We access publicly available data courtesy of Johns Hopkins University,
consisting of time-series data of recoveries, deaths, and confirmed cases stratified across several dozen groups (in this case, primarily geographic locations)~\cite{jhu-csse}. Our computations were performed on April 18, 2020. We denote cohorts or groups of cases by indices $g$, belonging to a set $G$. For example, $g$ could be `people under 60 years of age,' or `people in Wuhan.'
For time points $t = 1, 2, \ldots, T = 41$, we collect daily data as follows:
for each group $g \in G$ we collect $R_{t}^{g}$, $D_{t}^{g}$, and $C_{t}^{g}$, which correspond to the number of new recoveries, new deaths, and new cases reported on day $t$ within group $g$.
We drop the group superscript $g$ for population quantities:

\begin{equation}
R_{t} := \sum_{g \in G} R_t^g, \quad
D_{t} := \sum_{g \in G} D_t^g, \quad
C_{t} := \sum_{g \in G} C_t^g.
\end{equation}
\subsection{An Estimator Based on Dividing Deaths by Cases}
In early March 2020, the WHO estimate of the CFR, 3.4\% was widely reported~\cite{stelter_2020, who}. This estimate is obtained from a naive estimator;\footnote{To be clear, the exact form of their estimates is not made explicit in the WHO report.} 
specifically, the raw proportion of deaths among confirmed cases.
Formally, as of March 6, 2020,
\begin{equation}
\naive=\frac{\sum_{t}D_{t}}{\sum_{t}C_{t}} \approx 3.4\%.
\end{equation}
As of April 18, 2020, $\naive$ was $6.9\%$. However, as we establish in Appendix \ref{appendix-naivebiased}, in a setting without time delays, the naive CFR is asymptotically unbiased for the true CFR if and only if the probability of reporting is the same for fatal and nonfatal cases. Moreover, it is unbiased in finite samples if and only if reporting is perfect. As discussed in Section~\ref{sec:biases}, this is not true in the case of COVID-19. We also derive the finite-sample expectation of the estimator. Even asymptotically, the expectation of this estimator can become unboundedly far away from the true CFR as reporting goes to zero or the CFR goes to zero.

The naive estimator requires no complex modeling or tuning parameters and is easy to interpret. As we argued in Section~\ref{sec:biases}, there is no uniformly best method of measuring the CFR, and the naive estimator should be viewed as one in a constellation of estimators giving a heuristic idea of the causal CFR.  Nonetheless, the naive estimator can be improved at little cost, and indeed, in this work, we suggest applying a simple correction for time-dependent reporting rates and alleviate two problems with the naive estimator: time-lag between death and recovery, and time-dependence in the reporting rates of fatal and nonfatal cases.  

\subsection{An Estimator Based on Observed Outcomes} 
One can view the time lag in the numerator above (across the $\edge{D}{DF}$ link) as a consequence of `censoring' the data: a case has been identified, but the outcome is hidden. Methods for handling censored data have been studied for several decades in the statistical literature; in particular, in the context of the bootstrap~\cite{efron1981censored}. Although it is not the focus of our work, several others have already applied the bootstrap to COVID-19 data to find confidence intervals for other epidemiological parameters such as $R_0$~\cite{bootstrap1,gamma1}. This should also be done for the CFR for COVID-19, as \citet{bootstrapcfr} did for SARS, although the structure of the data used in that work differs from the current setting. 

There is also a very simple estimator that avoids censoring by using only observed data, namely:
\begin{equation}
  \observed = \frac{\sum_{t=1}^{T} D_t}{\sum_{t=1}^{T}D_t + \sum_{t=1}^{T} R_t} \approx 20.7\%.
\end{equation}  
The CFR calculated by this estimator is upwardly biased, and we will briefly discuss why.
This estimator accounts for the inflation of the denominator in the naive estimator via the relative time lag between $\edge{D}{RC}$ and $\edge{D}{DF}$. However, it assumes we observe the same fraction of recovered cases and fatal cases at the time of estimation. Thus, it has introduced a new bias, the relative reporting rate and time lag between $\edge{D}{DF}$ and $\edge{D}{DR}$. We formalize the asymptotic inferential target of this estimator in Appendix~\ref{appendix-obsbiased}. Note that in all cases, $\observed \geq \naive$. In fact, $\observed$ is exactly $3\naive$ on April 18th. This large discrepancy is due to under-reporting of recoveries, specifically within the United States, as we note in Section~\ref{sec:biases}. The United States has, as of April 19, roughly $40,000$ deaths and $70,000$ recoveries~\cite{jhu-csse}. Meanwhile, Spain has $20,000$ deaths and $80,000$ recoveries. Clearly, the reporting of recoveries in both nations is infrequent, and in the United States, it may be more than doubly so. The estimator $\observed$ illustrates the dangers of correcting one of many biases without considering total error. The estimator $\naive$ and the estimator we present in the next section do not use this recovery data.

\section{Likelihood Models}
\label{sec:models}
In this section, we describe a parametric model that, with respect to several strong modeling 
assumptions, accounts for two biases: time-varying reporting and disease-delayed mortality. 
For definitions and discussion of our model parameters, see Table~\ref{table:model-params}. 
With reference to Figure~\ref{graphicalmodel}, the model accounts for the 
time-dependence of $\edge{D}{DF}$ and from $\edge{D}{DR}$ (i.e., it models how the values of 
these conditional probabilities change as a function of time), and also for the 
time delay across those same edges.
This model was previously used by \citet{reich} for CFR estimation of 
influenza. It is a covariate-independent reporting model that assumes all nonfatal cases 
eventually recover, so it does not utilize the time series of recoveries.
Similar parametric models have been used for CFR 
estimation during other pandemics~\cite{timerequired, frome}.
When none of the biases in Section~\ref{sec:biases} other than the 
time dependence and time delay across $\edge{D}{DF}$ and $\edge{D}{DR}$ are large and the 
mathematical assumptions in the remainder of Section~\ref{sec:models} are satisfied, this estimator has a  
smaller total error than $\naive$ and $\observed$, evidenced by empirical evaluations in \citet{reich}.

\newcommand{\ra}[1]{\renewcommand{\arraystretch}{#1}}
\begin{table}
  \centering
  \begin{tabular}{@{}l p{0.9\textwidth}@{}}
    & \textbf{Definition} \\
    \midrule
      $\psi_{t, g}$ & Probability of diagnosis, given death from COVID-19, onset time $t$, and group $g$. \\
      $\phi_{t, g}$ & Probability of diagnosis, given recovery from COVID-19, onset time $t$, and group $g$. \\
    $p_{t, g}$ & Probability of death, given onset time $t$, within group $g$. \\
    $\eta_t$ & Probability of death $t$ days after onset, given death occurs. \\
  \end{tabular}
  \caption{Parameters and Variables Relevant to our Likelihood Models}
  \label{table:model-params}
\end{table}

Suppose that an individual is in group $g$ and has infection onset at time $\tonset$. 
Such a case has three possible outcomes, whose probabilities we define in Equation~\ref{defn:scenarioprobs}.
For further information, see also Table~\ref{table:scenarios}.
First, the individual may eventually recover and be diagnosed.
This occurs with probability $\recprob$, see Equation~\ref{eqn:rec}.
Secondly, they may eventually die, having been diagnosed. This occurs with probability $\deathprob$; see Equation~\ref{eqn:death}.
Finally, they may go entirely undiagnosed. This occurs with the remaining probability, $\undiagprob$; see Equation~\ref{eqn:undiag}.  In summary:
\begin{subequations}
  \label{defn:scenarioprobs}
  \begin{align}
    \recprob &= \phi_{\tonset, g} \left( 1 - p_{\tonset, g}\right),
    \label{eqn:rec}\\
    \deathprob &= \psi_{\tonset, g} p_{\tonset, g},
    \label{eqn:death}\\
    \undiagprob &= 1 - \rho^{(1)}_{\tonset, g} - \rho^{(2)}_{\tonset, g} =
    p_{\tonset, g}\left(1 - \psi_{\tonset, g}\right) +
    \left(1 - p_{\tonset, g})(1 - \phi_{\tonset, g}\right).  
    \label{eqn:undiag}
  \end{align}
\end{subequations}

Accordingly, at each onset time $\tonset$ and for each group $g$, there are $\recnum$, $\deathnum$, and $\undiagnum$ individuals who eventually recover, die, or go undiagnosed respectively.
Given a total number of cases within group $g$ with onset at time $\tonset$, denoted $\totalnum$, we model
the outcomes via a \emph{multinomial model}:
\begin{equation}\label{eqn:multinomial}
  (\recnum, \deathnum, \undiagnum) \simind \mathsf{Multinomial}\big(\totalnum, \recprob, \deathprob, \undiagprob\big),
  \quad \text{for all}~\tonset~\text{and}~g.
\end{equation}
We assume these are independent across onset times and group. Furthermore, we assume knowledge
of certain horizon probabilities. In order to define an estimator, we also need
probabilities $\eta_{t, \tonset, g}$, for $t \geq 0$. These are probabilities that, given an individual is in group
$g$ and has onset of infection at time $\tonset$, they die $t$ days later. We make the assumption that
$\eta_{t, \tonset, g} \equiv \eta_{t}$ for all $\tonset, g$. That is, these probabilities are time- and group-invariant.
See \citet{reich} for further analysis and evaluation of this model.

\begin{table}
  \centering
  \begin{tabular}{@{}l | ll@{}}
    & \textbf{Diagnosed} & \textbf{Undiagnosed} \\
    \midrule
    \textbf{Death} &  Included in scenario 1 & Included in scenario 3  \\
    \textbf{Recovery} & Included in scenario 2 & Included in scenario 3 \\
  \end{tabular}
  \caption{Outcome Scenarios for COVID-19 Patients With Onset at Time $\tonset$ and in Group $g$ \protect\footnotemark}
  \label{table:scenarios}
\end{table}

\footnotetext{ In our notation, scenario $i$ occurs with probability $\rho^{(i)}_{\tonset, g}$ and has count $N^{(i)}_{\tonset, g}$, for $i = 1, 2, 3$.}

Having stated the model, we now turn to the estimator. Let $N_{\tonset, g}$ denote the \emph{reported} total number of cases of
COVID-19 with onset at time $\tonset$ in group $g$. Unfortunately, this is not the quantity of true interest.
Instead, as mentioned above, we need $N^\ast_{\tonset, g}$, which is the number of number of both reported and \emph{unreported} cases.
Let $\E$ denote the expectation operator. In particular, $\E\nolimits_{N^\star_{\tonset, g}}$ will be an expectation with respect 
to the multinomial model in Equation~\ref{eqn:multinomial} indexed by $N^\star_{\tonset, g}$. 
As a simplifying assumption, we assume $\deathprob \equiv p_g$; that is, the group-specific death probability or CFR
is time-invariant. If the $p_g$ are small, then $N^\ast_{t, g} \approx N_{t, g}/\phi_{t, g}$, in which case from our
multinomial (Equation~\ref{eqn:multinomial}), it is easy to check that:
\begin{equation}
\E\nolimits_{N^\star_{\tonset, g}} \left[\deathnum \right] = \deathnum \deathprob \approx N_{\tonset, g} \frac{\deathprob}{\phi_{\tonset, g}}.
\end{equation}
In particular, if we assume that death is a rare event, then a Poisson approximation will be accurate:
\begin{equation}
\deathnum  \sim \mathsf{Poisson}\left(N_{\tonset, g} \frac{\deathprob}{\phi_{\tonset, g}}\right),
\end{equation}
where $N_{\tonset, g}$ denotes the number of cases with onset at time $\tonset$ within group $g$.
In view of Equation~\ref{eqn:death}, this may be rewritten as:
\begin{equation}
\deathnum  \sim \mathsf{Poisson}\left(N_{\tonset, g} \frac{\psi_{\tonset, g} p_{\tonset}}{\phi_{\tonset, g}}\right). 
\end{equation}
If we make either an assumption that the reporting rates are group-invariant, or that there is perfect fatal-case reporting, 
$\psi_{t, g} \propto \phi_{t, g}$, then it is possible to rewrite the model in the form:
\begin{equation}
    \E \left[\deathnum\right]  \approx N_{\tonset, g} + \beta_0 + \alpha_\tonset + \gamma_g,
\end{equation}
where $\beta_0$ is a proportionality constant, $\gamma_g$ is a group-specific parameter (the relative CFR),
and $\alpha_{\tonset}$ is a time-specific parameter. 
Finally, given these values, along with the death probabilities $\eta_t$, an expectation-maximization scheme
can be carried out to compute a maximum-likelihood estimator. 
For further details, see sections~3.2 and 3.3 of~\citet{reich}.
They show empirically that as long as $p_{g}$ stays below $0.05$, and their assumptions are approximately satisfied, the estimated CFR has relative error $<0.1$ as compared to the ground truth.
Their results also indicate that this model is insensitive to various misspecifications, including the distribution of deaths, $\eta_t$.
We confirm this in Figure~\ref{resultsfig} by sampling parameters of $\eta_t$
from their estimated confidence intervals~\cite{gamma1}.

\section{Results}
\label{sec:results}
\begin{figure}[t]
    \begin{subfigure}{\textwidth}
        \begin{minipage}[t]{0.33\textwidth}
            \includegraphics[width=\textwidth]{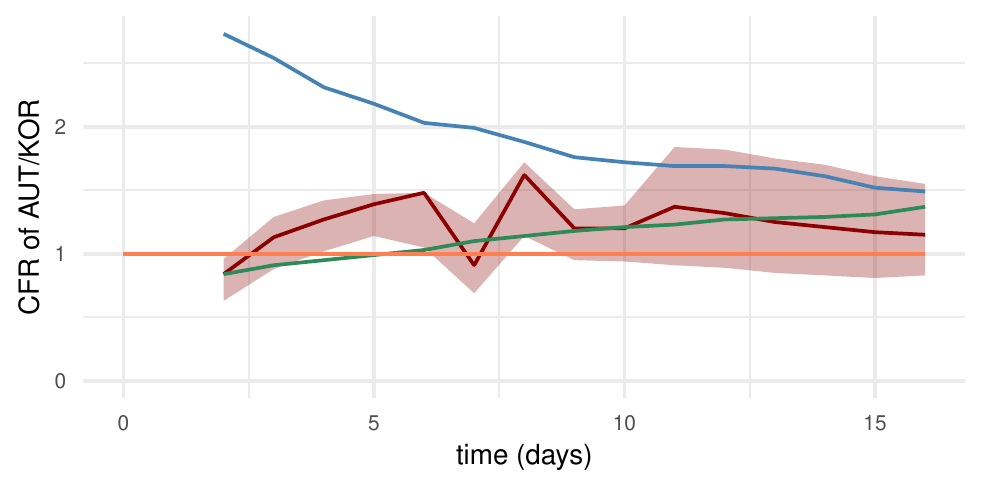}
        \end{minipage}
        \begin{minipage}[t]{0.33\textwidth}
            \includegraphics[width=\textwidth]{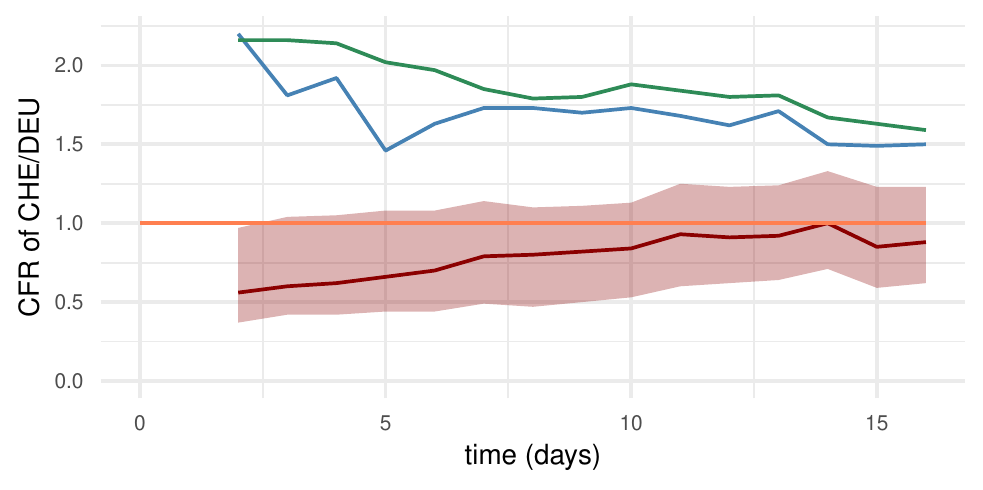}
        \end{minipage}
        \begin{minipage}[t]{0.33\textwidth}
            \includegraphics[width=\textwidth]{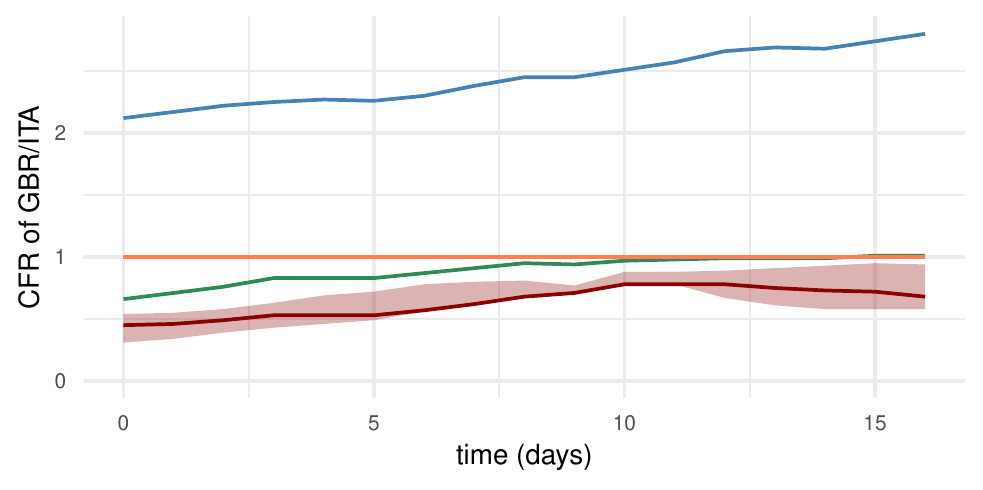}
        \end{minipage}
    \end{subfigure}
    \begin{subfigure}{\textwidth}
        \begin{minipage}[t]{0.33\textwidth}
            \includegraphics[width=\textwidth]{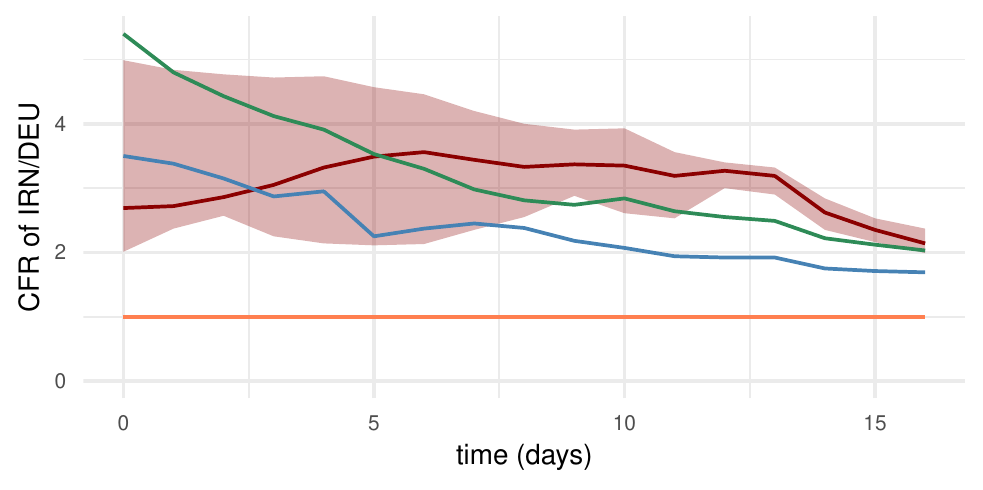}
        \end{minipage}
        \begin{minipage}[t]{0.33\textwidth}
            \includegraphics[width=\textwidth]{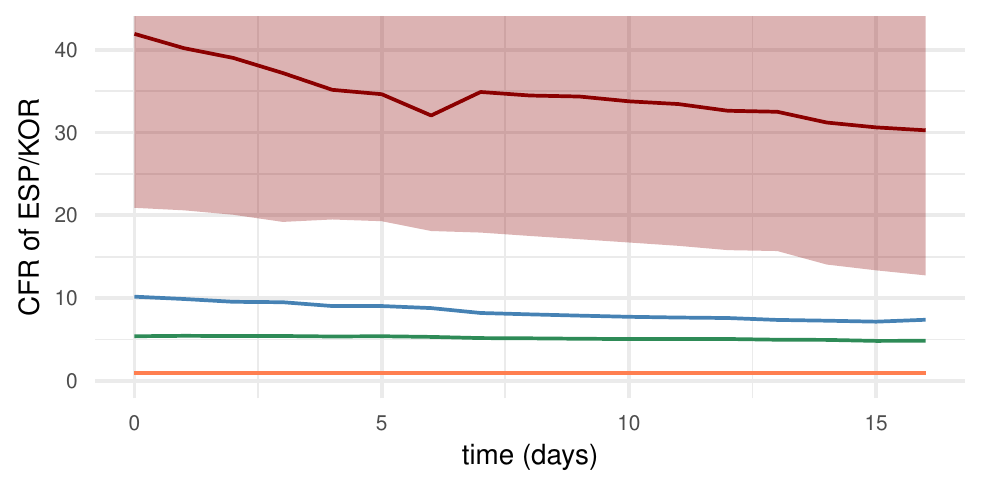}
        \end{minipage}
        \begin{minipage}[t]{0.33\textwidth}
            \includegraphics[width=\textwidth]{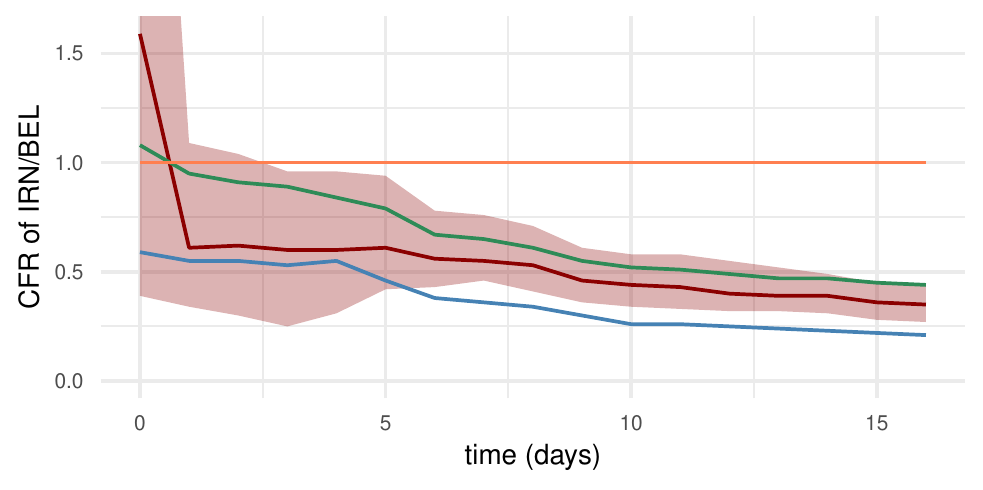}
        \end{minipage}
    \end{subfigure}
    \begin{subfigure}{\textwidth}
        \sisetup{
              table-align-uncertainty=true,
                separate-uncertainty=true,
            }
            \renewrobustcmd{\bfseries}{\fontseries{b}\selectfont}
            \renewrobustcmd{\boldmath}{}
        \centering
        \small
        \begin{tabular}{ @{}l c c c c c c@{} }
            
            \bfseries Estimator & \bfseries AUT/KOR & \bfseries CHE/DEU &\bfseries GBR/ITA &\bfseries IRN/DEU &\bfseries ESP/KOR &\bfseries IRN/BEL   \\
            \midrule
            \textcolor{OliveGreen}{$\boldsymbol{\naive}$} & $1.37$ & $1.59$ & $1.01$ & $2.03$ & $4.96$ & $0.44$  \\
            \textcolor{MidnightBlue}{$\boldsymbol{\observed}$} & $1.49$ & $1.5$ & $2.8$ & $1.69$ & $7.28$ & $0.21$ \\
            \textcolor{BrickRed}{$\boldsymbol{E_{\rm Reich}}$} & \thead{$1.15$ \\ $(0.83,1.55)$} & \thead{$0.81$\\$(0.62,1.23)$}&\thead{$0.68$ \\ $(0.58,0.94)$}
                                       & \thead{$2.03$ \\ $(1.99,2.37)$} & \thead{$30.27$ \\ $(12.75,55.23)$} &\thead{$0.38$ \\ $(0.27,0.44)$} \\
        \end{tabular}
    \end{subfigure}
    \caption{\textbf{The estimators $E_{\rm naive}$, $E_{\rm obs}$, and $E_{\rm Reich}$ presented
    as time series from April 2, 2020 to April 16, 2020.} Our estimator, in red, implements the
correction for time-dependent relative reporting rates between countries identified by their
ISO abbreviations. Sensitivity of our results to misparameterization of $\eta_t$ is reported 
by setting $\eta_t$ to be a discretized gamma distribution with mean $12.8-17.5$ and variance
$5.2-9.1$, the lower and upper extremes of the 95\% confidence intervals 
referenced~\protect\cite{gamma1}. The ribbon shows the maximum and minimum values of the estimator 
$E_{\rm Reich}$ at each time point under any combination of these conditions. The expectation 
maximization algorithm converged in all cases with negligible variance. We include the relative 
CFR of Spain to South Korea as an example of two countries for which our assumptions are 
particularly badly violated. Consequently, the method is unstable in that case (although we have 
no ground truth data for confirmation). Notice each plot has a different vertical axis scaling. 
We have included an orange line at a relative CFR of $1$ to indicate the point when two countries 
have the same estimated CFR; this provides a reference point between the plots. } 
    \label{resultsfig}
\end{figure}
We report the results of our analysis of open-sourced COVID-19 data from Johns Hopkins, under the assumption that the reporting rates $\psi_{t}$ and $\phi_{t}$ are group-invariant. We contribute an open-source multithreaded implementation of \citet{reich} and a plotting utility that will allow reproducibility of these results, as shown in Figure~\ref{resultsfig}. Finally, we report the relative CFR of women to men in Germany and Belgium using sex-disaggregated data from \citet{riffe}.

\subsection{Estimates of Relative CFRs}
\begin{figure}
  \centering
  \includegraphics[width=0.6\linewidth]{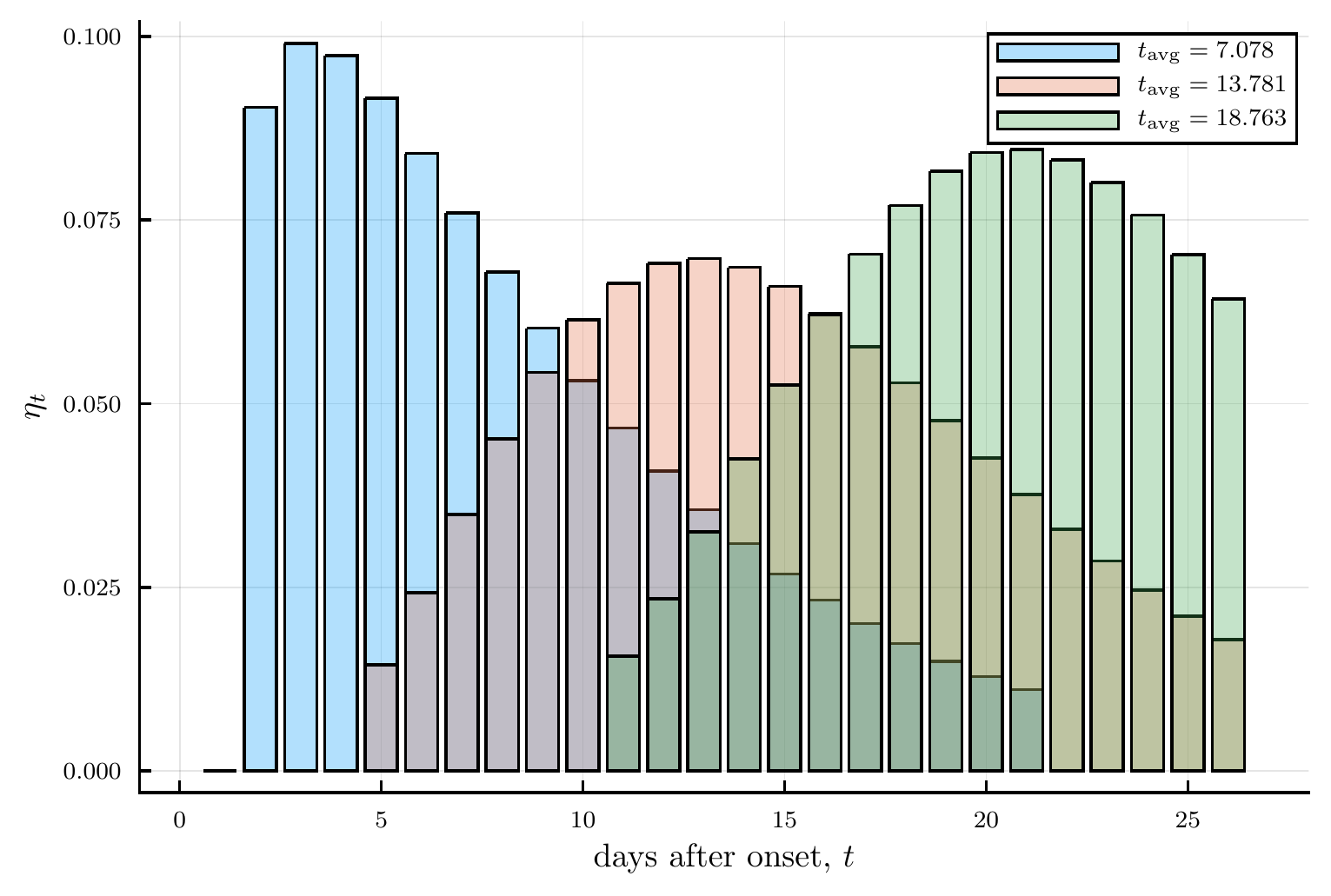}
  \caption{\textbf{Illustration of distribution of death times for fatal cases, $\mathbf{\eta_t}$, in days post onset.}
              Our method assumes knowledge of the probability of death for a fatal case $t$ days post-onset of COVID-19.
          This data was estimated by fitting a gamma distribution to the fatality time horizons from Chinese data in early February,
          2020~\protect\cite{gamma1}. We discretize the distribution by day and also truncate it to $25$ days long, both for numerical stability and also
      because very few deaths occurred past this point in the real data. The mean time to death was $15.0$ (95\% CI $12.8-17.5$). The standard deviation was $6.9$ (95\% CI $5.2-9.1$), which we also used in our sensitivity analysis above. The three separate gamma distributions plotted above have different choices of mean value for illustrative purposes, to show the qualitative effect of changing the parameter.}
  \label{fig:plotdist}
\end{figure}

The corrected relative CFRs, calculated for six combinations of nations, are listed in Figure~\ref{resultsfig}. In some cases, such as the comparison between England (GBR) and Italy (ITA), our estimator flips the direction of the relative CFR. In other words, $\naive$ and $\observed$ suggest that England has a higher CFR than Italy, while $E_{\rm Reich}$ suggests otherwise. The same effect happens in the case of Switzerland (CHE) vs. Germany (DEU), with an additional shrinkage in the distance toward $1$, indicating the relative CFR is more similar than $\naive$ and $\observed$ would suggest. The estimate of the relative CFR for Spain (ESP) to South Korea (KOR) predicted by $E_{\rm Reich}$ is high at $30.27$. Although we assumed in Section~\ref{sec:models} that the relative CFR is constant in time, we report our results as a time series in Figure~\ref{resultsfig}. We obtain this time series by calculating results as if we had run our estimator on every day from April 2, 2020, and April 16, 2020, using the cumulative data.

Using the data from \citet{riffe}, we calculated the relative CFR of women to men in Germany and Belgium. In Germany, $\naive=1.51$ and $E_{\rm Reich} = 1.14$ (Sensitivity $1.14-1.22$). In Belgium, $\naive=1.68$ and $E_{\rm Reich} = 1.25$ (Sensitivity $1.13-1.26$). Time-series data of recoveries is not available, so we could not calculate $\observed$. We chose Germany and Belgium because the data from these nations had about two months of seemingly reliable, day-by-day, sex-disambiguated data that roughly matched the numbers from Johns Hopkins. The dataset from Riffe was still under development at the time we ran these estimates.

\subsection{Choosing $\eta_t$}\label{sec:exp}
As described in Section~\ref{sec:models}, we assume access to probabilities $\eta_t$ that indicate the probability of death occurring for a fatal case $t$ days post-onset of COVID-19.
Since our model indexes time by day, we need to set $\eta_t$ for integers $t \geq 0$.
Our choice of distribution is the best-fit discretized gamma distribution to the fatality time horizons from Chinese data (shape parameter $k = 4.726$ and scale $\theta = 3.174$)~\cite{gamma1}.
These parameters were roughly consistent across 
several other studies~\cite{gamma2,meantimedeath}. We discretized the probability density function $\eta_t$ to the
days $t =0, \dots, \final$. 
Formally, after selecting a mean parameter $\deathavg > 0$, we determine the
probabilities $\eta_t$ by
\footnote{The notation $\propto$ is hiding a positive normalization constant to make $\eta$ a probability measure.}
\begin{equation}\label{defn:deathdist}
  \eta_t \propto t^{k-1} \e^{-t/\theta},
  \quad t = 0, \dots, \final.
\end{equation}

Stated differently, for a given mean parameter $\deathavg$, we define a probability measure $\eta \in \R^\final_+$, on
$t \in \{0, \dots, \final\}$, according to Equation~\ref{defn:deathdist}.
See Figure~\ref{fig:plotdist} for an illustration of this distribution.
In our experiments, we truncate at $\final=25$, both for numerical stability and also because very few deaths occur after $25$ days in the data used to fit the gamma distribution.

\section{Discussion}
\label{sec:discussion}
We emphasize again that the procedure that we have presented for estimation of relative CFR seeks to address only a subset of 
the biases that impinge upon the ascertainment of this important population-level parameter.
We explicitly account for the time-dependence of reporting rates that may differ between fatal 
and nonfatal cases. We have separate time-dependent reporting rates for cases that will eventually be fatal 
or nonfatal, addressing the fact that reporting is higher among severe cases. Deaths are known to vary with some combination 
of health care quality and age, which can be quantified with a relative CFR estimate. 
Our covariate-independent reporting rate assumption likely does not hold in practice. Indeed, the relative CFR of Spain with respect to Korea (two countries  
whose time-dependent reporting rates are probably different) yields a value of $30.27$, likely speaking to the limitations of this method, although we do not have ground-truth. Although~\citet{reich} present extensive experimental evaluations and some theory indicating that the method outperforms
$\naive$ under given modeling assumptions, it is not generally possible to check how closely these
assumptions hold, due to overparameterization of the unrestricted model. This issue may be mitigated by working 
with domain experts who understand each group's sampling and reporting patterns. Another issue is that our estimator uses parameters $\eta_t$ that are not estimated strictly from surveillance data but rather from individualized death times~\cite{gamma1}. 
\clearpage
\begin{wrapfigure}{r}{0.5\textwidth}
    \includegraphics[width=0.5\textwidth]{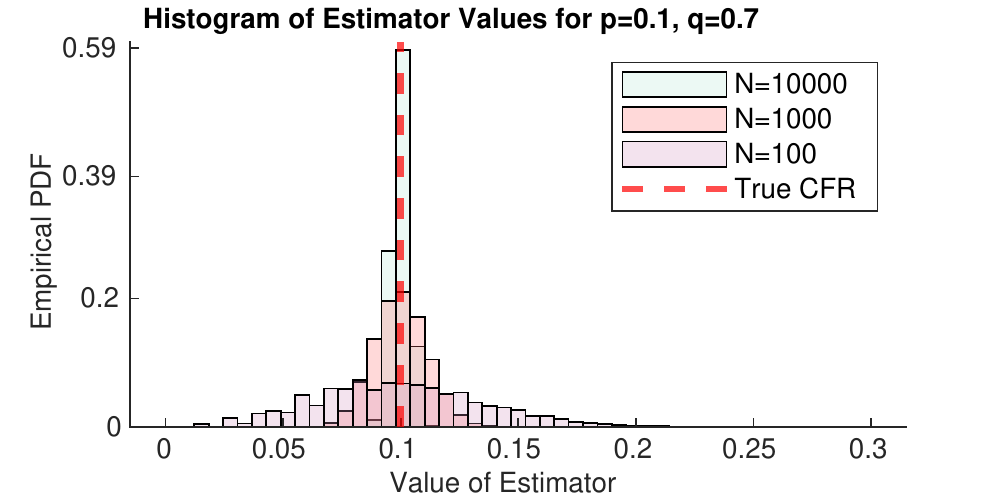}
    \includegraphics[width=0.5\textwidth]{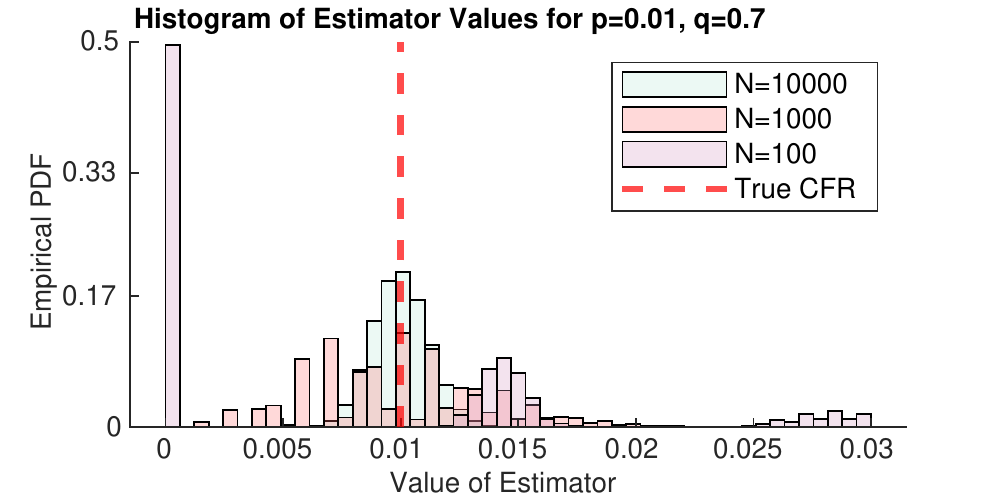}
    \vspace{5mm}
    \includegraphics[width=0.5\textwidth]{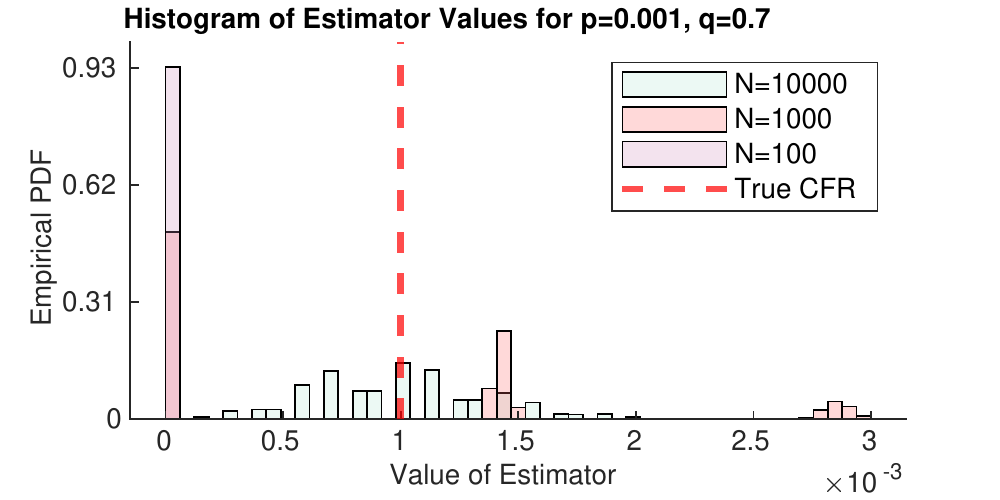}
    \centering
    \resizebox{0.43\columnwidth}{!}{%
    \begin{tabular}{ @{}l|c c c @{} }
        \sisetup{
              table-align-uncertainty=true,
                separate-uncertainty=true,
            }
            \renewrobustcmd{\bfseries}{\fontseries{b}\selectfont}
            \renewrobustcmd{\boldmath}{}
        & \bfseries \textdelta/\textit{p}=0.1 & \bfseries \textdelta/\textit{p}=0.01 & \bfseries \textdelta/\textit{p}=0.001 \\
        \midrule
        $\mathbf{q=0.9}$ & $N=1$ & $N=2$ & $N=3$ \\
        $\mathbf{q=0.5}$ & $N=4$ & $N=7$ & $N=10$ \\
        $\mathbf{q=0.1}$ & $N=22$ & $N=44$ & $N=66$ \\
    \end{tabular}
}
    \vspace{5mm}
    \caption{\textbf{Under the assumptions in the contact-tracing scenario, $\boldsymbol{\naive}$ converges after a small number of samples to a nearly unbiased estimator of the CFR.} The bound on the number of samples $N$ derived in Appendix~\ref{appendix-naivebiased}, $N\geq \Bigl\lceil\tfrac{\log(\delta/p)}{\log(1-q)}\Bigr\rceil$ (Equation~\ref{eq:naivefinite}) was used to calculate the values in the table. Notice $N$ is a function of $\delta/p$ (the acceptable relative error) and $q$ (the reporting rate). The empirical distribution functions of $\naive$ with different parameters of $p$ and $N$ and a reporting rate $q=0.7$ are plotted. Notice that as $p$ decreases, detecting a case will require larger $N$.}
    \label{contactfigure}
    \vspace{-4mm}
\end{wrapfigure}

We believe that the maximum-likelihood estimator that we have presented may provide a more 
valid correction of relative reporting rates between German women and men rather than between South Korean people and Italian people, given that 
reporting rates by sex may be closer to identical than reporting rates by country, although 
biases by sex still exist~\cite{sexbias}.
Demographers have argued that releasing 
    data stratified by sex, age, and other demographic groups would aid in understanding the spread and fatality rates of COVID-19~\cite{demographic}. Although certain teams like \citet{riffe} are currently assembling this data, many agencies are reporting such strata infrequently or not at all, making data collection difficult. To our knowledge, there is no well-established data repository (like the Johns Hopkins repository) that contains time-series data of deaths and cases stratified by sex, age, and so on.

Many of the key biases that we reviewed in Section~\ref{sec:biases} remain unaddressed in
current data-collection and data-analysis pipelines.  Variations in the nature of the
population within the sampling frame that gets tested, due to government- or geography-specific protocols,
will cause any CFR estimate to be unreliable. In particular, details in the definitions of terms across countries 
and times can result in severe bias in time-series data; for example, China's explicit policy 
was that they would not report asymptomatic cases until April 1, 2020, when the policy changed~\cite{asymptomaticchina}. Accounting for many of the biases we have discussed may be possible with great effort by many data analysts. 
However, it is equally important for the statistical community to channel much of its energy into a unison
clarion call to governments: to obtain estimates to support consequential policy-making, we need more and better data. 

Contact tracing is a particularly powerful way to obtain data that allow otherwise intractable biases to be controlled, since it expands the sampling frame to include a much larger portion of our target population, specifically mild cases.
Contact tracing is the process of reaching out to all 
individuals (`contacts') who were recently exposed to a known 
COVID-19-positive individual, removing them from circulation, and monitoring 
their health. The same is done for contacts of contacts, and so on, for an 
appropriate number of iterations. We suggest that all of these contacts 
should be tested for COVID-19 one incubation period after exposure, regardless 
of whether or not they are symptomatic. The number of data points gleaned from 
this strategy will be lower than the number of data points from surveillance data. However, 
the population sampled using this strategy would be closer to the target population, 
since it would include asymptomatic cases. Furthermore, there is no issue with time lag, 
since these cases can be tracked systematically. Specifically, assume the nonresponse 
rate to contact tracing is identical for asymptomatic and symptomatic cases. As we 
prove in Appendix~\ref{appendix-naivebiased}, this is the exact condition under 
which $\naive$ is an asymptotically unbiased estimator. Moreover, the estimator
has desirable finite-sample properties in such a setting. Letting $p$ be the true CFR and $q$ be the 
reporting rate among infected cases, we have that $\naive$ lies within a range $\delta$ of $p$ in 
$N=\Bigl\lceil\tfrac{\log(\delta/p)}{\log(1-q)}\Bigr\rceil$ samples; see Equation~\ref{eq:naivefinite} below. As seen in Figure~\ref{contactfigure}, 
$N$ does not need to be too large to insure that the bias of $\naive$ is small, although with small $p$, sampling any fatal cases requires $N$ to be on the order of $1/p$ in this simplified model.

Contact tracing does not eliminate all biases. The assumption that nonresponse rates do 
not vary by case severity will not hold unless responses are mandatory, possibly introducing 
significant error, especially as $p$ becomes small. Assay sensitivity 
and specificity may still cause errors. Most importantly, care must be taken to make valid 
inferences about the desired target population based on individualized contact-tracing data 
that may come from a restricted sample. One major hurdle is estimation of $p$ when it is small: 
as shown in Figure~\ref{contactfigure}, if death is a very rare event, $N$ would need to be 
large in order to ensure enough fatal cases are sampled. Finally, such data may be easier 
to collect and release in some countries and jurisdictions than others. For example, within 
the United States, medical privacy and consent laws may make it difficult to ever test a 
truly random sample of the population, or to release the fine-grained data necessary for 
corrected estimators. These 
challenges, outside the scope of our work, are well studied in the field of survey sampling and reweighting. 

\subsection*{Disclosure Statement}
The authors have no conflicts of interest to declare.

\subsection*{Acknowledgments}
A.\ N.\ A.\ was partially supported by the National Science Foundation Graduate Research Fellowship Program.
R.\ P.\ was partially supported by a UC Berkeley University Fellowship via the ARCS Foundation.
We wish to thank Constance Angelopoulos for illustrating Figure~\ref{graphicalmodel}, Esther Rolf and Ilija Radosavovic for reading and commenting on the manuscript, \href{https://hypernetwork.io/}{Hypernet} for providing compute resources, and Anthony Ebert for contributing code and comments to our GitHub pre-release (\href{https://github.com/AnthonyEbert/COVID19data}{see his Git at https://github.com/AnthonyEbert/COVID19data}). Finally, we thank the editor and reviewers of the \textit{Harvard Data Science Review} for providing valuable feedback.
 
\subsection*{Contributions}
We use the \href{https://www.elsevier.com/authors/journal-authors/policies-and-ethics/credit-author-statement}{CRediT taxonomy of contributions}. A.\ N.\ A.\: conceptualization, methodology, software, formal analysis, experiments, data curation, original draft, editing, visualization. R.\ P.\: conceptualization, methodology, formal analysis, editing, visualization. R.\ V.\: editing. M.\ I.\ J.\: conceptualization, resources, editing, supervision.
\begin{appendices}
\numberwithin{equation}{section}

\setcounter{equation}{0}

\section{Derivation of the Expectation of $\naive$}
\label{appendix-naivebiased}
In this section we derive the expectation of $\naive$.  Our derivation will employ a stripped-down notation, since here we deal with individual random variables for each COVID-19-infected person instead of time-series data.  
Index the infected population with the integers $\{1,\cdots,N\}$. Let $T_i \sim {\rm Ber}(p)$ be a Bernoulli random variable representing whether or not person $i\in\{1,\dots,N\}$ died, and let $W_i \sim {\rm Ber}(q)$ be a Bernoulli random variable representing whether or not person $i$ was diagnosed with the virus. We want to estimate $p$, but we only have the reported number of deceased patients with COVID-19, $V_i=T_iW_i$, $i\in\{1,\dots,N\}$. Defining $r \defeq {\rm Cov}(T_i,W_i)$, the joint distribution of $(T_i,W_i)$ can be expressed as a contingency table:

\begin{table*}[h!]
\centering
\begin{tabular}{ @{}l|c c @{} }
    &$\mathbf{T_i=1}$ & $\mathbf{T_i=0}$ \\
    \midrule
    $\mathbf{W_i=1}$ & $r+pq$ & $q(1-p)-r$  \\
    $\mathbf{W_i=0}$ & $p(1-q)-r$ & $1+r-p-q+pq$  \\
\end{tabular}
\end{table*}

Several of the results discussed in the main article follow as simple consequences of the calculation of the expectation of $\naive$: (1)$\naive$ is unbiased for $p$ in finite samples if and only if $q=1$; (2) $\naive$ is unbiased for $p$ as $N\to \infty$ if and only if $r=0$; (3) if there is an $\epsilon>0$ error in the estimation of $r$, for example due to incorrect attribution of fatalities to COVID-19, then $\naive$ has unbounded expectation $q\to 0$ and unbounded relative error as $p\to 0$; and (4) if $r=0$, the smallest $N$ such that $|\mathbb{E}\left[\naive\right]-p|\leq \delta$ is $N=\Bigl\lceil\tfrac{\log(\delta/p)}{\log(1-q)}\Bigr\rceil$.

The distribution of $V_i$ is Bernoulli with $\mathbf{P}\left[V_i = 1\right] = r+pq$. Also define $\gamma_1=\mathbb{P}\left[W_i=1 \; \middle |\; T_i=1\right]=(r+pq)/p$. Applying the tower property of conditional expectation and using the exchangeability of the $(T_i, W_i)$ pairs, we have:
\begin{equation}
    \mathbb{E}\left[\frac{\sum_{i=1}^{N}V_i}{\sum_{j=1}^{N}W_i}\right] = \sum_{i=1}^{N}\mathbb{E}\left[\frac{V_1}{\sum_{j=1}^{N}W_j}\right]=N\mathbb{E}\Bigg[T_1\mathbb{E}\bigg[W_1\frac{1}{W_1+\sum_{l=2}^{N}W_l} \;\bigg| \; T_1\bigg]\Bigg].
\end{equation}
Since $W_1$ is independent of $W_{2, \ldots, N}$, we can express the sum in the denominator as a binomial random variable, $B\sim {\rm Bin}(N-1,q)$, since it is a sum of $N-1$ i.i.d. Bernoulli random variables ($W_2, \dots, W_N$) with parameter $q$. Note the fact that $\mathbb{E}\left[\frac{1}{1+B}\right] = ((1-(1-q)^N))/Nq$. Then, evaluating the innermost expectation first:
\begin{equation}
    N\mathbb{E}\Bigg[T_1\mathbb{E}\bigg[W_1\frac{1}{W_1+B} \;\bigg|\; T_1\bigg]\Bigg]=
    N\mathbb{E}\Bigg[T_1\gamma_1\mathbb{E}\bigg[\frac{1}{1+B}\bigg] \;\Bigg|\; T_1\Bigg]=
    p\frac{\gamma_1}{q}(1-(1-q)^N).
\end{equation}
Finally, substituting for $\gamma_1$, we obtain the final form:
\begin{equation}
    \label{eq:naiveexp}
    \mathbb{E}\left[\naive\right]=\frac{r+pq}{q}(1-(1-q)^N).
\end{equation}
Recall that $q$ is the probability of reporting given an infection, and $p$ is the probability of death given an infection. Since $1-q=0$ implies $r=0$ because $W$ becomes deterministic, $\naive$ is unbiased for $p$
if and only if $q=1$. Moreover, if $q\neq1$, taking $N\to \infty$ shows $\naive$ is asymptotically biased for $p$ if and only if $r=0$. Both of these conditions are violated for any real disease. Interestingly this empirical CFR is not constrained to be an underestimate, and can overestimate $p$ if $p\leq\tfrac{r(1-(1-q)^N)}{q(1-q)^N}$.

Under the assumption $r\geq\epsilon$, the (asymptotic) overestimate can be unboundedly bad. This may arise if there is an $\epsilon$ error in estimating the covariance (which we assume to be nonnegative), because some people are diagnosed with COVID-19 but their death is not \textit{caused} by COVID-19. In this context:
\begin{equation}
\label{eq:naiveinf}
\lim_{q \to 0}  \frac{r + pq}{q} \geq\lim_{q\to0} \frac{\epsilon+pq}{q}=\infty.
\end{equation}
In other words, as the rate of reporting, $q$, decreases or the covariance between death and reporting increases, the CFR estimate gets worse, ultimately becoming infinitely bad, as long as there is a spuriously positive relationship between death and diagnosis. Similarly, the ratio $\naive/p$ can become infinitely bad as the product $pq$ decreases. Note that if there is no spurious relationship, $r\to 0$ as $p\to 0$ or $q\to 0$ since $W$ and $T$ become deterministic under those conditions. In the case of COVID-19, neither $q$ nor $p$ are near zero, but the limiting case helps to exhibit the qualitative performance of the estimator---it becomes more bias-prone with smaller $p$ and $q$ and with larger $r$.

Finally, assume that $W$ and $T$ are independent, so $r=0$. Then, $|\mathbb{E}\left[\naive\right] - p|\leq\delta$ implies that
    $p(1-q)^N\leq\delta,\mbox{ and by some simple algebra, } N\geq\frac{\log(\delta/p)}{\log(1-q)}$.
Constraining $N$ to be the smallest $N\in\mathbb{N}$ such that $|\mathbb{E}\left[\naive\right] - p|\leq\delta$ gives 
\begin{equation}
    \label{eq:naivefinite}
    N=\Biggl\lceil\frac{\log(\delta/p)}{\log(1-q)}\Biggr\rceil.
\end{equation}

\section{Derivation of the Asymptotics of $\observed$}
\label{appendix-obsbiased}
We borrow notation and proof technique from \citet{reich}. We use the same notation as the main article. This proof applies to the group-independent reporting-rate model. In Reich et al., a similar proof is shown that applies in the case of the constant-proportion assumption. 

In addition to the notation in the main article, define: $d_{t,g}\defeq\mathbb{E}[D_{t,g}]=N^*_{t,g}p_g\psi_t$ and $r_{t,g}\defeq\mathbb{E}[R_{t,g}]=N^*_{t,g}(1-p_g)\phi_t$. Also, introduce two nonrandom functions, $F_d : \R_+ \to [0,1]$ and $F_r: \R_+ \to [0,1]$, where $F_d(t)$ represents the fraction of confirmed, fatal cases who have died by time $t$. Similarly $F_r(t)$ represents the fraction of confirmed, nonfatal cases who have recovered by time $t$. During an active outbreak, we have $F_d < 1$ and $F_r < 1$. Finally, define $T$ as the current time; all sums over time below have an upper limit of $T$ unless otherwise specified. We seek an asymptotic limit for:
\begin{equation}
\observed =\frac{F_d(T)\Sigma_{t_2}D_{t_2,g}}{(F_d(T)\Sigma_{t_1}D_{t_1,g})+(F_r(T)\Sigma_{t_1}R_{t_1,g})}.
\end{equation}

By the weak law of large numbers, $D_{t,g}$ and $R_{t,g}$ converge to their expectations, so we have:
\begin{equation}
    \frac{F_d(T)D_{t,g}}{N^*_{t,g}} \overset{p}{\to} \frac{F_d(T)d_{t,g}}{N^*_{t,g}} = F_d(T)p_g\psi_t,
\end{equation}
and similarly,
\begin{equation}
    \frac{F_r(T)R_{t,g}}{N^*_{t,g}} \overset{p}{\to} \frac{F_r(T)r_{t,g}}{N^*_{t,g}}= F_r(T)(1-p_g)\phi_t.
\end{equation}

Now we focus on the denominator. We have to introduce a ``smoothness'' assumption: the number of infected people at each timestep, $N^*_{t_1,g}$, has a constant ratio with respect to the number of infected people at each other timestep, $N^*_{t_2,g}$. In particular, $\lambda_{t_1,t_2,g}$ corresponds roughly to the growth rate of the disease. Although this quantity would vary based on many factors in a real setting, we assume it to be a constant here. In particular, as $N^*_{t_1,g} \to \infty$ and $N^*_{t_2,g} \to \infty$, 
\begin{equation}
\frac{N^*_{t_1,g}}{N^*_{t_2,g}} \to \lambda_{t_1,t_2,g}.
\end{equation}
Therefore, we have by Slutsky's theorem that:
\begin{equation}
    \frac{F_d(T)D_{t_1,g}+F_r(T)R_{t_1,g}}{N^*_{t_2,g}} \overset{p}{\to} \lambda_{t_1,t_2,g}(F_d(T)p_g\psi_{t_1}+F_r(T)(1-p_g)\phi_{t_1}).
\end{equation}

Now, applying the weak law of large numbers and our assumption:
\begin{equation}
    \frac{F_d(T)D_{t_1,g}+F_r(T)R_{t_1,g}}{N^*_{t_2,g}} = \frac{N^*_{t_1,g}}{N^*_{t_2,g}}(\frac{F_d(T)D_{t_1,g}}{N^*_{t_1,g}} +\frac{F_r(T)R_{t_1,g}}{N^*_{t_1,g}}) \overset{p}{\to} \lambda_{t_1,t_2,g}(F_d(T)p_g\psi_{t_1}+F_r(T)(1-p_g)\phi_{t_1}).
\end{equation}
Then, by Slutsky's theorem on the sum which is the denominator of $\observed$ divided by $N^*_{t_2,g}$,
\begin{equation}
    \Sigma_{t_1}\left[\frac{F_d(T)D_{t_1,g}+F_r(T)R_{t_1,g}}{N^*_{t_2,g}}\right] \overset{p}{\to} \Sigma_{t_1}\lambda_{t_1,t_2,g}(F_d(T)p_g\psi_{t_1}+F_r(T)(1-p_g)\phi_{t_1}).
\end{equation}

Now, considering one term from the sum over $t_2$ in $\observed$, we have:
\begin{align}
  \frac{F_d(T)D_{t_2,g}}{\Sigma_{t_1}F_d(T)D_{t_1,g}+F_r(T)R_{t_1,g}}&=\frac{F_d(T)D_{t_2,g}/N^*_{t_2,g}}{\Sigma_{t_1}(F_d(T)D_{t_1,g}+F_r(T)R_{t_1,g})/N^*_{t_2,g}}\\
                                            &\overset{p}{\to} \frac{F_d(T)p_g\psi_{t_2}}{\Sigma_{t_1}(\lambda_{t_1,t_2,g}(F_d(T)p_g\psi_{t_1} + F_r(T)(1-p_g)\phi_{t_1}))}.
\end{align}
Finally, rearranging terms and appealing once more to Slutsky's theorem:
\begin{equation}
\observed \overset{p}{\to} \Sigma_{t_2}\frac{F_d(T)p_g\psi_t}{F_d(T)(\Sigma_{t_1}\lambda_{t_1,t_2,g}p_g\psi_t)+F_r(T)(\Sigma_{t_1}\lambda_{t_1,t_2,g}(1-p_g)\phi_t)}.
\end{equation}
This is clearly a biased estimator of $p_g$.
\end{appendices}

\bibliographystyle{apalike}
\bibliography{references}

\end{document}